\documentclass[prb,preprint,preprintnumbers,amsmath,amssymb,showpacs,superscriptaddress]{revtex4-1}
\usepackage{amssymb}

\usepackage{graphicx}
\usepackage{amsmath}
\usepackage{bm}
\usepackage{color}
\usepackage[normalem]{ulem}

\begin{document}

\title{Observation of A$_g^1$ Raman mode splitting in few layers black phosphorus encapsulated with hexagonal boron nitride}

\author{J.\ M.\ Urban}\thanks{These authors contributed equally to the work}
\affiliation{Laboratoire National des Champs Magn\'etiques Intenses, UPR 3228, CNRS-UGA-UPS-INSA, Grenoble and Toulouse, France}
\author{M.\ Baranowski}\thanks{These authors contributed equally to the work}
\affiliation{Laboratoire National des Champs Magn\'etiques Intenses,
UPR 3228, CNRS-UGA-UPS-INSA, Grenoble and Toulouse, France}
\affiliation{Department of Experimental Physics, Faculty of
Fundamental Problems of Technology, Wroclaw University of Science
and Technology, Wroclaw, Poland}
\author{A.\ Surrente}
\affiliation{Laboratoire National des Champs Magn\'etiques Intenses, UPR 3228, CNRS-UGA-UPS-INSA, Grenoble and Toulouse, France}
\author{D.\ Wlodarczyk}
\affiliation{Institute of Physics, Polish Academy of Sciences, al.\ Lotnik{\'o}w 32/46, 02-668 Warsaw, Poland}
\author{A.\ Suchocki}
\affiliation{Institute of Physics, Polish Academy of Sciences, al.\ Lotnik{\'o}w 32/46, 02-668 Warsaw, Poland}
\author{G.\ Long}
\affiliation{Physics Department and the Center for Quantum
Materials, the Hong Kong University of Science and Technology, Hong
Kong, China}
\author{Y.\ Wang}
\affiliation{Physics Department and the Center for Quantum
Materials, the Hong Kong University of Science and Technology, Hong
Kong, China}
\author{L.\ Klopotowski}
\affiliation{Institute of Physics, Polish Academy of Sciences, al.\ Lotnik{\'o}w 32/46, 02-668 Warsaw, Poland}
\author{N.\ Wang}
\affiliation{Physics Department and the Center for Quantum
Materials, the Hong Kong University of Science and Technology, Hong
Kong, China}
\author{D. K.\ Maude}
\affiliation{Laboratoire National des Champs Magn\'etiques Intenses, UPR 3228, CNRS-UGA-UPS-INSA, Grenoble and Toulouse, France}
\author{P.\ Plochocka}\email{paulina.plochocka@lncmi.cnrs.fr}
\affiliation{Laboratoire National des Champs Magn\'etiques Intenses, UPR 3228, CNRS-UGA-UPS-INSA, Grenoble and Toulouse, France}


\date{\today}

\begin{abstract}

We investigate the impact of the encapsulation with hexagonal boron nitride (h-BN) on the Raman spectrum of few layer black phosphorus. The encapsulation results in a significant reduction of the line width of the Raman modes of black phosphorus, due to a reduced phonon scattering rate. We observe a so far elusive peak in the Raman spectra $\sim$4cm$^{-1}$ above the A$_{\text{g}}^1$ mode in trilayer and thicker flakes, which had not been observed experimentally.  The newly observed mode originates from the strong black phosphorus inter-layer interaction, which induces a hardening of the surface atoms vibration with respect to the corresponding modes of the inner layers. The observation of this mode suggests a significant impact of h-BN encapsulation on the properties of black phosphorus and can serve as an indicator of the quality of its surface.

\end{abstract}

\maketitle

Black phosphorus (BP), the most stable allotropic form of phosphorus, has been synthesized more than a century ago
\cite{bridgman1914two}. Its recent rediscovery \cite{ling2015renaissance} is related to the possibility of exfoliating thin
layers, adding BP to the rapidly emerging family of two-dimensional (2D) materials. Its unique physical properties
\cite{carvalho2016phosphorene,liu2015semiconducting} make it attractive for the next generation of electronic
\cite{na2014few,luo2014temporal,li2014black, liu2014phosphorene}, optoelectronic \cite{engel2014black,youngblood2015waveguide}
and thermo-electronic applications \cite{fei2014enhanced, luo2015anisotropic,jang2015anisotropic}. Similarly to other layered
semiconductors, such as transition metal dichalcogenides (TMDs) or graphene, the physical properties of BP can be tuned by
varying the number of layers \cite{tran2014layer}. What distinguishes BP from other 2D semiconductors is the simultaneous
presence of strong in-plane anisotropy, and a relatively strong inter-layer coupling.

In single layer BP the atoms form a puckered honeycomb lattice, giving rise to a strong in plane anisotropy of the optical
\cite{qiao2014high,wang2015highly}, electrical \cite{li2014black,xia2014rediscovering} and thermal \cite{fei2014enhanced,
luo2015anisotropic,jang2015anisotropic} properties. Each phosphorus atom forms covalent bonds with its three closest neighbors
leaving a pair of lone electrons with elongated orbitals  perpendicular to the plane \cite{luo2015large, hu2016interlayer}. This
results in a significant overlap of the electronic wave function between the layers \cite{hu2016interlayer}. The inter-layer
coupling is therefore much stronger than in the case of other layered semiconductors such as TMDs, hexagonal boron nitride (h-BN)
or few layer graphene\cite{carvalho2016phosphorene, castellanos2014isolation,lu2014plasma}. Due to the strength of the
inter-layer coupling, the band gap can be tuned over a wide range of energies from 0.3\,eV for a bulk crystal up to more than
2\,eV for monolayer black phosphorus \cite{Zhang2017, Asahina, Kim723, Morita1986, liang2014electronic,surrente2016excitons,surrente2016onset}. In
contrast to most TMDs, the gap is always direct even for bulk crystals \cite{liang2014electronic,Li2017}. The lattice dynamics
are also influenced by the strong inter-layer interaction. In contrast to weakly coupled TMDs or few layer graphene
\cite{song2016physical,lu2014plasma, ferrari2006raman}, in BP the vibrational modes are either unaffected or soften (red shift) with an
increasing number of layers \cite{lu2014plasma,hu2016interlayer}. This trend is opposite to the predictions of the classical
model of coupled harmonic oscillators \cite{wieting1972interlayer}, indicating that the inter-layer interaction in BP is of a
different nature than the weak van der Waals forces \cite{luo2015large,hu2016interlayer}. Moreover the strong inter-layer interaction is predicted to manifest in different out of plane vibration frequencies of surface and inner layers\cite{hu2016interlayer}.

Despite the interesting electronic properties, there are a number of technological challenges to overcome if BP is to be used in
applications. Notably, BP is highly sensitive to the ambient conditions. The lone electron pairs are responsible for the rapid
degradation of BP in air \cite{koenig2014electric,castellanos2014isolation,wood2014effective,alsaffar2017raman}. The surface
readily reacts with oxygen and moisture, leading to the formation of oxygen defects\cite{ziletti2015phosphorene, ziletti2015oxygen, utt2015intrinsic, yang2015interpreting, island2015environmental, edmonds2015creating, doganov2015transport, han2016strongly, wood2014effective, carvalho2016phosphorene, Favron2015} (which can modify the doping level of BP), and ultimately to the formation of oxide and phosphoric acid \cite{Favron2015}. This reactivity is one of the main hindrances for
the production of functional devices. Encapsulation in different materials such as PMMA \cite{alsaffar2017raman,tayari2015two},
epitaxial organic monolayers \cite{Zhaoorganic}, h-BN \cite{cao2015quality,doganov2015transport}, and Al$_2$O$_3$
\cite{luo2014temporal,na2014few, wood2014effective} has been used in an attempt to stabilize BP. Few layer h-BN is
particularly promising for encapsulation, as h-BN encapsulated BP becomes resistant to oxidation and exhibits excellent electrical
properties \cite{chen2015high, long2016achieving, avsar2015air}. However, very little is known about the optical properties of encapsulated BP. Since encapsulation is an inevitable step in the production of stable BP based devices, it is crucial to fully understand its impact on the electronic and optical properties of BP.

In this work we present a systematic study of few layer BP encapsulated in h-BN using Raman spectroscopy. This method has
previously been applied in multiple studies of non-encapsulated BP \cite{castellanos2014isolation, lu2014plasma,
Favron2015,phaneuf2016polarization}. We find that encapsulation with h-BN leads to significant and unexpected narrowing of the
Raman lines, accompanied by the appearance of a so far elusive Raman mode. This Raman peak appears slightly above the
A$^1_\text{g}$ mode and is most pronounced in the trilayer. We interpret this observation as a consequence of the different
vibrational frequency of atoms in the layer at the surface resulting from the covalent nature of interlayer bondings. This effect was predicted theoretically \cite{hu2016interlayer} but
has not been observed experimentally so far. Our findings indicate that Raman spectroscopy provides a measure of the BP
crystalline quality. Since the photoluminescence emission vanishes in h-BN/BP/h-BN heterostructures\cite{cao2015quality}, Raman
spectroscopy is the only readily available optical technique that can be used for samples characterization/selection before
further processing. Moreover, the observed surface mode gives a direct handle to the condition of the surface \emph{i.e.} its oxidation
state and opens the way for the investigation of the interaction of the encapsulating material with the surface of BP.

\begin{figure}[h!]
\centering
\includegraphics[width=0.9\linewidth]{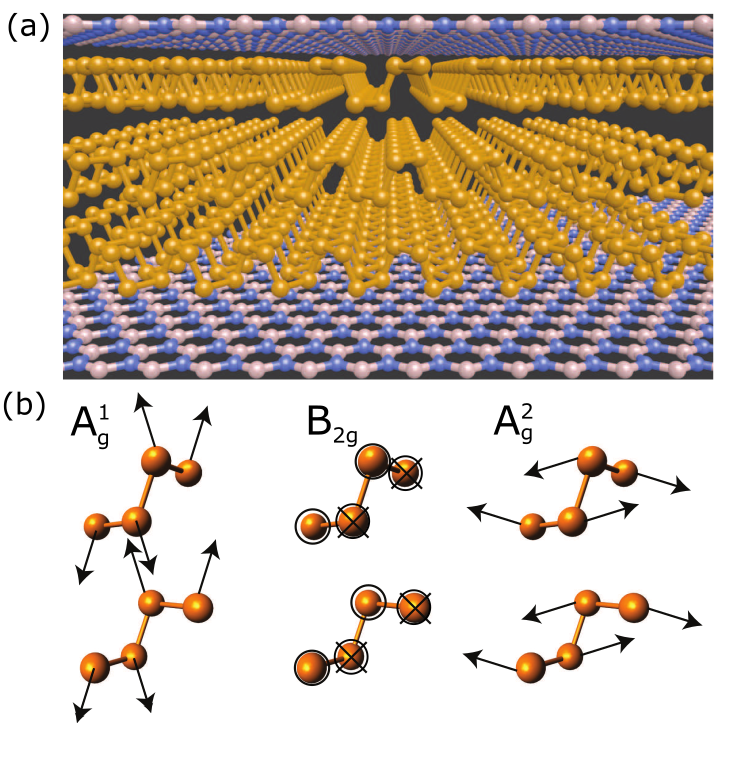}
\caption{(a) Crystal structure of black phosphorus encapsulated in h-BN. (b) Schematic showing the Raman-active modes
A$^1_\text{g}$, B$_{2\text{g}}$ and A$^2_\text{g}$.} \label{fig:scheme}
\end{figure}

For our investigation, several flakes of BP with different number of layers were encapsulated in hexagonal boron nitride. A
schematic illustration of such a structure is shown in Fig.\,\ref{fig:scheme}(a). Thin flakes of BP are prepared by
micro-mechanical exfoliation of a single crystalline BP on silicon substrates covered with 300\,nm-thick SiO$_{2}$. Bottom and
top hexagonal boron nitride (h-BN) sheets are prepared on another silicon substrate and a PMMA thin film, respectively. The top
h-BN sheet is then applied to pick up the BP flake from the silicon substrate. The formed h-BN/BP structure is then transferred
to the top surface of the bottom h-BN sheet to form h-BN/BP/h-BN heterostructure. Due to the sensitivity of BP flakes to air, the
exfoliation and transfer processes are performed in a glove box filled with highly pure nitrogen gas to protect the BP flakes
from oxidation and degradation. The heterostructure is then stabilized by annealing at $250 - 350^{\circ}$C in Ar atmosphere for
15 hours. Annealing improves the contact between h-BN and BP, removing small bubbles and adsorbants that might be trapped between the heterolayers during the fabrication process \cite{avsar2015air}, and it reduces charge trap density\cite{avsar2015air,chen2015high}. Raman measurements were performed under ambient conditions at room temperature using 630\,$\mu$W of a 532\,nm
laser. The light was focused down to a spot approximately 1\,$\mu$m in diameter using a $100\times$ objective (N.A.=0.9). The
spectral resolution of the setup was 0.5 cm$^{-1}$. Flakes were mapped over the areas of several $\mu$m  with approximately
1\,$\mu$m step. We did not observe any degradation of the samples during the measurements over a period of a few weeks.

The unit cell of BP contains 4 atoms, which results in 12 vibrational modes at the $\Gamma$ point of the Brillouin zone. Six of the vibrational modes are Raman active in bulk and few layer BP (few layers and bulk BP share the same character table since
they are described by the same point group $D_{2h}$): three acoustic modes (B$_{\text{3g}}^1$, B$_{\text{1g}}$,
B$_{\text{3g}}^2$) and three optical modes (A$_{\text{g}}^1$, B$_{\text{2g}}$, A$_{\text{g}}^2$)\cite{carvalho2016phosphorene}.
Here we focus our Raman investigation on the three most intense optical vibrational modes, presented schematically in Fig.\,\ref{fig:scheme}(b).

\begin{figure}[h!]
\centering
\includegraphics[width=1.0\linewidth]{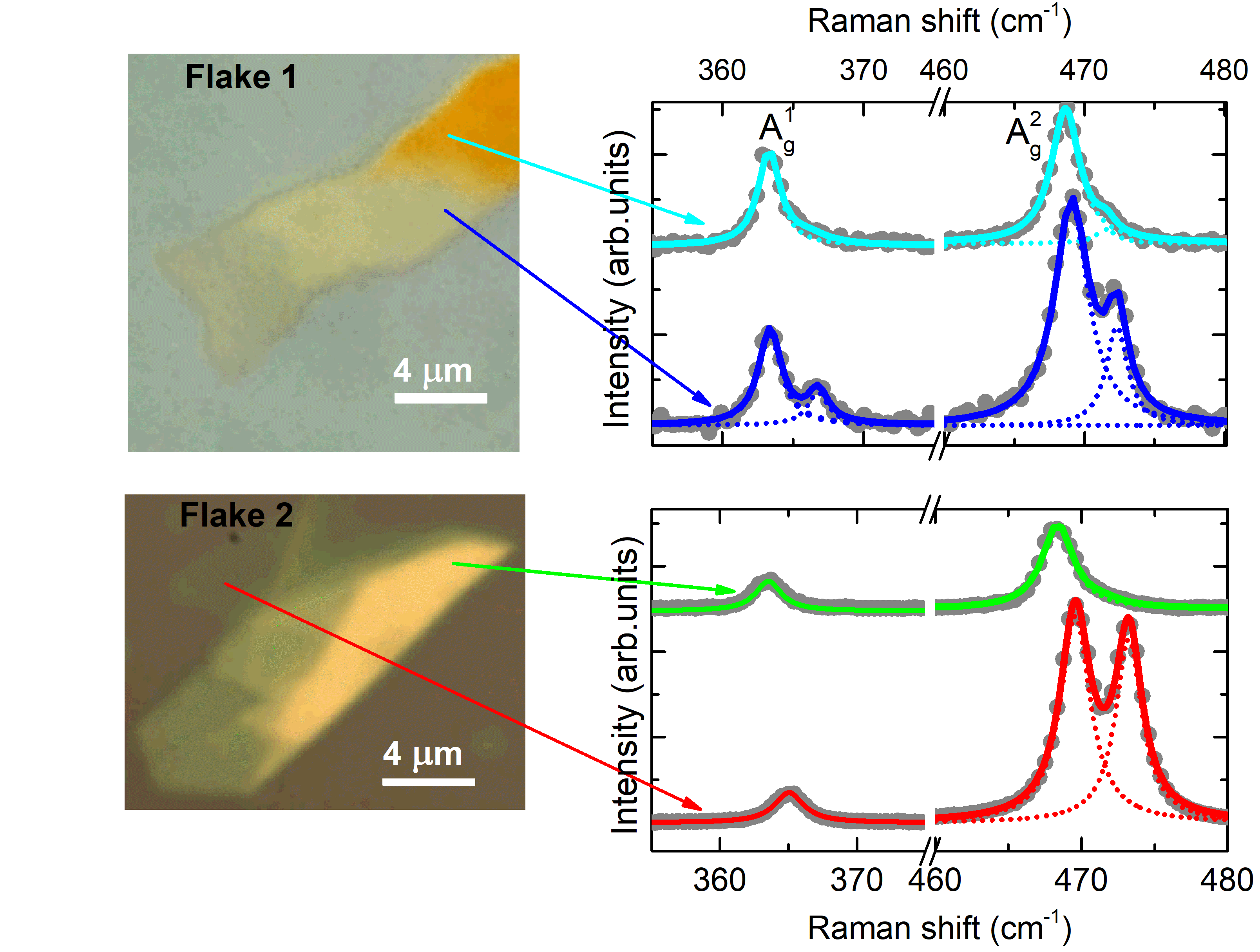}
\caption{Optical images of two of the investigated black phosphorus flakes without (flake 1) and with (flake 2) h-BN
encapsulation. The Raman spectra (grey symbols) are obtained on two different positions of each flake as indicated. The solid
lines are fitted using a single or double Lorentzian line shape.} \label{fig:optical}
\end{figure}

Fig.\,\ref{fig:optical} shows optical images of two BP flakes together with representative Raman spectra acquired
at different positions on the flake. In the optical image we can distinguish regions with different thickness (interference
effects lead to a different color depending on the number of layers). For flake 1 the optical image was
taken before encapsulation. We stress that all Raman data are taken on h-BN encapsulated flakes. Raman peaks related to
A$^1_\text{g}$, A$^2_\text{g}$ and B$_{2\text{g}}$ vibrational modes are observed for all flakes. While the Raman spectrum of the
B$_{2\text{g}}$ mode does not change significantly from place to place (see supplementary information), the shape of the A$^1_\text{g}$ and A$^2_\text{g}$ Raman peaks clearly depends on the flakes thickness. Moreover, for some places on the sample additional
peaks above A$^1_\text{g}$ and A$^2_\text{g}$ are clearly visible. While the peak on the high energy side of A$^2_\text{g}$ mode
was observed previously \cite{Favron2015,phaneuf2016polarization}, to the best of our knowledge this is the first experimental observation of an additional
Raman mode a few wave numbers above A$^1_\text{g}$.

\begin{figure}[h!]
\centering
\includegraphics[width=1.0\linewidth]{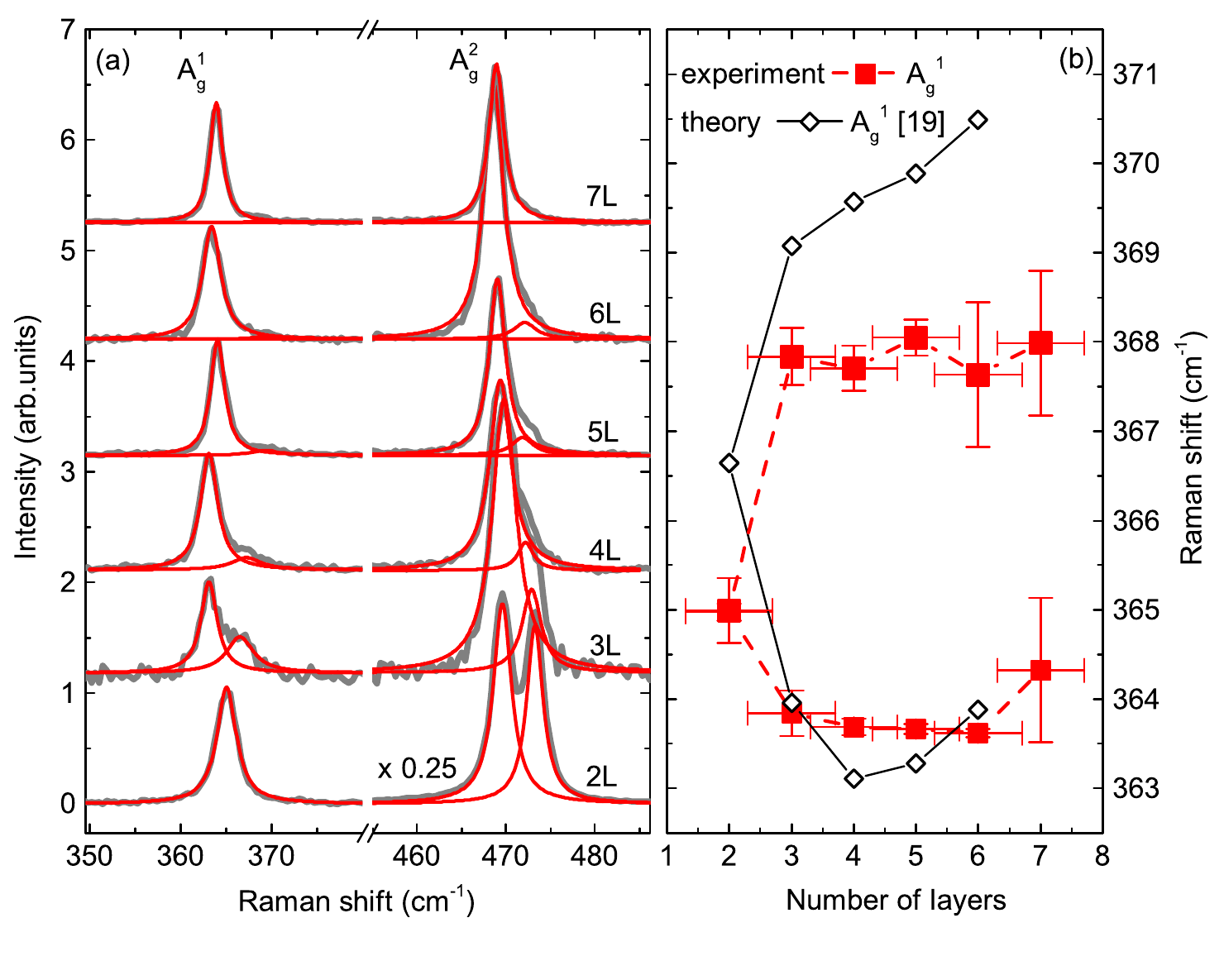}
\caption{(a) Evolution of A$^1_\text{g}$ and A$^2_\text{g}$ peaks as a function of the number of layers (grey)
and together with Lorentzian fits (red). The spectra are normalized by the A$^1_\text{g}$ peak intensity and the A$^2_\text{g}$
peak has been rescaled for clarity for the bilayer spectrum. All spectra have been measured on flake 3, with the exception of the
2L spectrum measured on flake 2.  (b) Averaged (over all spectra measured at different places) dependence of
A$^1_\text{g}$ inner and surface mode energy as a function of the number of layers (red squares) together with the predictions of Ref. \onlinecite{hu2016interlayer} (open diamonds). \label{fig:raman_evolution}}
\end{figure}

\begin{figure}[t]
\centering
\includegraphics[width=0.9\linewidth]{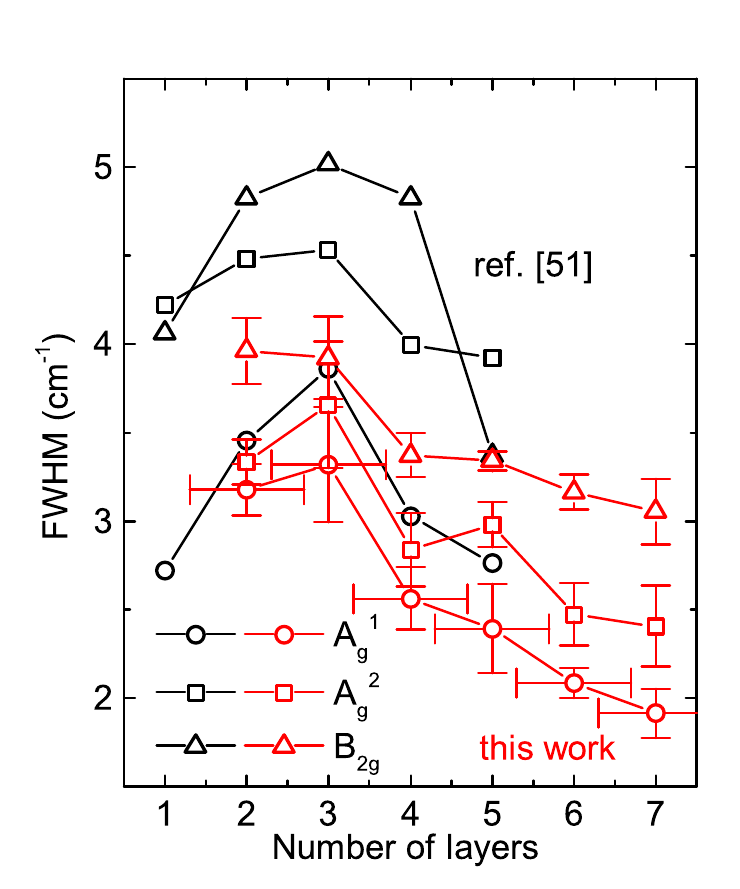}
\caption{Comparison of the Raman mode FWHM of BP encapsulated with hexagonal boron nitride (red) and bare flakes
(black) taken after Ref.\ \onlinecite{phaneuf2016polarization}.}
\label{fig:FWHM}
\end{figure}

To reliably extract the position, width and intensity of the Raman modes we fit the Raman peaks with either a single or double
Lorentzian line shape as presented in Fig.\,\ref{fig:optical} The error bars presented in Fig.\ \ref{fig:raman_evolution} and \ref{fig:FWHM} represent the standard deviations of the distributions of the peak positions and line widths obtained on all the positions of the investigated flakes. To understand the evolution of the Raman peaks we estimate the
number of layers involved. A precise determination of the BP thickness is not straightforward since the layers are encapsulated
by h-BN, making AFM measurements difficult to interpret. Therefore, we use the relative shift of the A$^2_\text{g}$--B$_{2\text{g}}$ as the indicator of layer thickness\,\cite{lu2014plasma} (detailed discussion of this approach is presented in the
supplementary information). Having determined the number of layers, we can follow the evolution of the A$^2_\text{g}$ and
A$^1_\text{g}$ Raman peaks as a function of the flake thickness. Figure \ref{fig:raman_evolution} shows representative Raman
spectra for 2--7 layers measured on flake 2 and flake 3 (only flake 2 had a bilayer region). A strong characteristic Raman line
$\simeq 4$\,cm$^{-1}$ above the A$^2_\text{g}$ mode is observed especially for bilayer BP. The intensity of this line systematically decreases with an increasing number of BP layers. This additional Raman mode has been already reported in the literature \cite{phaneuf2016polarization, Favron2015}. It is related to the doubling of the number of atoms in the bilayer unit cell, which allows the Davydov conversion of two out-of-phase IR-active mode B$_{2\text{u}}$\cite{sugai1985raman, phaneuf2016polarization} modes to a Raman allowed mode with energy very close to that of the A$^2_\text{g}$ mode. This infrared converted mode is particularly robust since
the high energy peak can still be distinguished even in the case of 5 layers. This is most probably related to the reduced line
width of our Raman peaks after encapsulation of the BP with hexagonal boron nitride.

The main result of this work is the first experimental observation of an additional, easily distinguishable Raman mode, which appears as a peak shifted by $\sim4$\,cm$^{-1}$ on the high wave number side with respect to the A$^1_\text{g}$ mode for three layer and thicker BP. The intensity of this peak is
highest for the trilayer, and it decreases rapidly with increasing number of layers. A similar behavior of the Raman
modes (splitting and rapid decrease of the intensity with the number of layers) has been observed in other layered
materials \cite{song2016physical,staiger2015splitting, phaneuf2016polarization,Golasa2017}. It has been attributed to Davydov
splitting, induced by van der Waals interaction between the layers, which gives rise to an additional force felt by atoms
vibrating out of phase in neighboring layers. However, the nature of the inter-layer interaction is fundamentally different in BP. Theoretical calculations
predict a hybridization of the lone electron pairs in adjacent layers, which leads to a large charge density in the region between the neighboring layers \cite{luo2015large,hu2016interlayer}. The inter-layer interactions in black phosphorus are thus of quasi-covalent character\cite{luo2015large,hu2016interlayer}. The Davydov-related splitting should result in a different number of peaks depending on the number of interacting BP layers. In contrast, we always observe only two peaks with relative intensities monotonically changing as a function of the number of layers \cite{song2016physical}. Moreover, the
splitting observed here is much larger than the full width half maximum of the A$^1_\text{g}$ mode of previous studies
\cite{Favron2015,phaneuf2016polarization} (see also Fig.\ \ref{fig:FWHM}) for non-encapsulated BP. Therefore, if the splitting of the
A$^1_\text{g}$ mode was due to the out of phase vibration of adjacent layers, it should have been easily observed in the previous works, which is not the case. Based on these arguments and on the good qualitative agreement between the observed splitting and the theoretical predictions \cite{hu2016interlayer} (see Fig.\ \ref{fig:raman_evolution}(b)), we attribute the additional peak to the splitting of the A$^1_\text{g}$ mode into surface and inner modes, which can be observed in our samples due to their superior quality. A particularly important role in this respect is played by the protection of surface layers from oxidation and by the narrowing of the Raman lines induced by the h-BN capping layers, as will be discussed further. 

 \textit{Ab initio} calculations \cite{hu2016interlayer} show that strong and highly directional interlayer interactions arising from the
electronic hybridization of the lone electron-pairs lead to anomalous surface phenomenon where the surface layer is stiffer than the
inner ones. Thus, the atoms at the surface of BP vibrate with higher frequency than those inside. The predicted splitting between the
inner and surface modes appears for 3 and more layers exactly as observed in our data. The splitting is predicted to be
$\simeq1$\,cm$^{-1}$ for B$_{2\text{g}}$ and $\simeq 6$cm$^{-1}$ for A$^1_\text{g}$ \cite{hu2016interlayer}. The predicted
splitting of the B$_{2\text{g}}$ mode is smaller than the Raman line width, so only the splitting of the A$^1_\text{g}$ mode
is observed experimentally. Our data are in good qualitative agreement with the theoretical predictions \cite{hu2016interlayer}, as demonstrated by Fig.\ \ref{fig:raman_evolution}
(b), with the experimental shift of the
high energy peak slightly smaller than theoretically predicted ($\sim$1--2cm$^{-1}$). The smaller energy of both the surface mode and the A$^1_\text{g}$ mode of bilayers as compared to theoretical predictions may originate from interactions of the surface layers with the capping h-BN (the calculations were performed for BP suspended in vacuum \cite{hu2016interlayer}). Due to the lone electron pairs of BP, a significant interaction with encapsulating layers can be expected for BP. Theoretical calculations predict a significant redistribution of the carriers across BP/BN and BP/graphene interfaces \cite{cai2015electronic} and BP/Al$_2$O$_3$ interface \cite{sun2017first}, which can affect the electronic properties of BP, as demonstrated by the expected increase of the BP bandgap in the case of graphene-encapsulated BP \cite{Padilha}. Moreover, in BP encapsulated with h-BN the photoluminescence is quenched \cite{cao2015quality}. These results suggest that the encapsulation with h-BN affects the electronic properties of BP. Therefore, small discrepancies between the theory and our measurements of the A$^1_\text{g}$ splitting might be a hallmark of the interaction between the outer layers of BP and h-BN.

The surface origin of the high energy Raman mode can also explain why it is not observed in uncapped BP. It is well established
that the Raman signal is quenched in oxidized BP \cite{Favron2015,alsaffar2017raman}. Since the oxidation proceeds layer by layer
the signal from the surface layers will be quenched while the signal from the inner layers remains strong. Encapsulation
stabilizes the surface preserving the Raman modes associated with the vibration of surface atoms.  Nevertheless,
with increasing thickness the bulk response becomes dominant (due to the volume) and the Raman mode from the surface is not
observed (its amplitude is less than 2--3 times the noise level) for more than 6--7 layers.

The h-BN encapsulation affects also the line width of the A$^1_\text{g}$, A$^2_\text{g}$ and B$_{2\text{g}}$ peaks, as shown in Fig.\,\ref{fig:FWHM}. All Raman lines measured on encapsulated samples are narrower, as demonstrated by the comparison of our data (averaged over all the spectra acquired on 4 flakes) with the literature data. We can see that the FWHM of A$^2_\text{g}$ and B$_{2\text{g}}$ modes measured on bare BP \cite{phaneuf2016polarization} is $\sim$1cm$^{-1}$ larger than that we measured on our encapsulated samples. We observe a similar effect for the inner A$^1_\text{g}$ mode, which in the case of our encapsulated samples is $\sim$0.5cm$^{-1}$ narrower.

The reduction of FWHM is most probably related to
the reduction of the phonon boundary and impurity scattering rate\cite{sun2013spectroscopic}, due to the better quality of
BP protected by boron nitride. It is interesting to note that in non encapsulated BP, the broadening of the
peaks does not change significantly with time (over several days) under ambient conditions \cite{Favron2015,alsaffar2017raman}.
This suggests that the broadening is mainly related to the layers adjacent to the oxidized layers. Since the oxidation occurs
layer by layer there is always one layer next to the oxidized layer, which explains why the FWHM remains constant despite the
continued degradation of the uncapped BP. The boron nitride encapsulation performed in  nitrogen atmosphere prevents the initial
degradation of the first layer and adsorption processes, thereby reducing the Raman line width with respect to uncapped black phosphorus.

To summarize, Raman spectroscopy has been used to investigate the vibrational properties of few-layer BP flakes encapsulated
with hexagonal boron nitride as a function of the number of layers. A previously unobserved observed splitting of the A$^1_\text{g}$
Raman mode has been demonstrated for trilayer and thicker flakes. The origin of the splitting is attributed to different
vibration frequency of the inner and surface layers \cite{hu2016interlayer}. This effect results from the strong inter-layer
interaction present in BP. The encapsulation of BP is crucial for the observed splitting since h-BN prevents surface layer
oxidation. The positive impact of encapsulation on the BP layers quality is confirmed by the narrow lines of the Raman peaks.

\section*{Author contributions}
DKM and PP coordinated the project. GL and YW fabricated the samples under NW supervision. JMU, DW, A.\ Suchocki and LK performed Raman spectroscopy measurements. JMU, MB, A.\ Surrente, LK, DKM and PP analyzed the data and discussed the results. JMU, MB, AS, LK, DKM, and PP wrote the manuscript with input of all other co-authors.

\section*{Conflicts of interest}
There are no conflicts of interest to declare.

\begin{acknowledgments}
This work was partially supported by ANR JCJC project milliPICS, the R{\'e}gion Midi-Pyr\'en\'ees under contract MESR 13053031,
the BLAPHENE project under IDEX program Emergence and by ``Programme Investissements d'Avenir'' under the program
ANR-11-IDEX-0002-02, reference ANR-10-LABX-0037-NEXT and by the PAN--CNRS collaboration withing the PICS 2016-2018 agreement. D.
Wlodarczyk and A. Suchocki were supported by the National Science Center of Poland grant DEC-2012/07/B/ST5/02080.
\end{acknowledgments}

\bibliographystyle{rsc}
\bibliography{BibBlackP}

\end{document}